\documentclass[
,final            
,numberedheadings 
  ]
  {aipproc}
\layoutstyle{6x9}

\newcommand{\be}{\begin{equation}}
\newcommand{\ee}{\end{equation}}
\newcommand{\ba}{\begin{eqnarray}}
\newcommand{\ea}{\end{eqnarray}}

\begin{document}

\title{The scalar sector  in the Myers-Pospelov model}


\author{C. M. Reyes, L.  Urrutia and J. D. Vergara}{
  address={Instituto de Ciencias Nucleares \\ Universidad Nacional Aut\'onma de M\'exico\\
  Circuito Exterior, C.U.\\04510 M\'exico, D. F.}
}

\begin{abstract}
 We construct  a perturbative expansion of the scalar sector in the
 Myers-Pospelov model, up to second order in the Lorentz violating parameter and  taking into account its
 higher-order time derivative character. This expansion allows us to construct an hermitian positive-definite
 Hamiltonian which provides a correct basis  for quantization. Demanding that the modified normal frequencies remain real
 requires the introduction of an upper bound in the magnitude $|\mathbf{k}|$ of the momentum, which is a manifestation of the effective character of the model.  The free scalar propagator, including the corresponding modified dispersion relations, is also calculated to the given order, thus providing the starting point to consider  radiative corrections when interactions are introduced.
\end{abstract}

\maketitle

\section{Introduction}

The basic aim of this and previous work \cite{RUV1} is to define a
consistent quantization of the Myers-Pospelov (MP) model considered as an
effective field theory which provides \textbf{perturbative }
corrections to standard QED. This model incorporates Lorentz
invariance violating (LIV) corrections to QED, which are codified
by dimension five operators which are C, T  conserving but P
violating. Our main concern with the construction, which we expect
to play a major role in defining the appropriate way of making sense
of the effective model, is that we recover the Lorentz invariant
results of standard QED when such dimension five operators are
turned off. In other words, we are looking for a smooth
interpolation procedure between a Lorentz violating description and
a Lorentz invariant one for QED. This point of view has been
successfully carried on in Ref. \cite{ALFARO}, where LIV is codified
by a dimensionless parameter modifying the integration measure in
momentum space appearing in the calculation of one-loop processes.
In this case LIV arises only from radiative corrections associated
to standard particle Lagrangians; that is to say the zeroth order
approximation is just the Standard Model of particles. The fact that
LIV should be treated as a perturbation over the standard dynamics
is supported by the very stringent limits which terrestrial
experiments together with astrophysical observations set upon the
parameters which label such violation. This requirement poses
additional interesting challenges to this problem because we need
 to deal with higher-order time derivatives (HOTD) theories. The
perturbative treatment of these, over the normal lower-order time
derivative cases, is a non-trivial task, but fortunately it is well
described in the literature
\cite{JAEN,SIMON,ELIEZER,CHENGETAL,CHENGETAL1}. As an additional
motivation to deal with quantum corrections in this model we mention
the fine tuning problems recently reported \cite{GV}.

In this work we will concentrate in the scalar sector of the model,
that can be used as the charged matter realization of scalar
electrodynamics,  which will be considered in a future work. Preliminary descriptions
of the photonic and
fermionic sector were already given in Ref. \cite{RUV1}.
 We will set up the correct zeroth order quantization,
identifying the new propagating normal modes described by their
modified dispersion relations (MDR)  up to second order in the LIV parameter.
We will also present the corresponding propagator, with the final aim to calculate radiative corrections which are
quadratical in the LIV parameters.

\section{Perturbative approach to higher order time derivative theories}

The general method for dealing with the canonical  description of
HOTD theories was given a long time ago in \ Ref.
\cite{OSTROGRADSKI}. In order to highlight some general features
of these theories we briefly review their basic properties in a
non-degenerate  mechanical setting
$\left(\frac{{\partial}^2L}{\partial q^{(k)}\partial
q^{(k)}}\right)\neq 0$,\ where the fields depend only upon the
time coordinate and
\begin{equation}
q^{(k)}(t)=\frac{d^{k}q(t)}{dt^{k}}.
\end{equation}
The generalization incorporating additional
coordinates $q(t)\rightarrow q_I(t),\,  I=1, \dots, N$ is direct
and the introduction of space dependent  fields goes along similar
lines as in the standard transition from mechanics to field
theory.

If the highest time derivative in the Lagrangian is of order $k$ , $%
L=L(q(t),...,q^{(k)}(t))$,
the corresponding phase space will be of dimension $2k$, been
characterized
by$\;k\;$coordinates$:Q_{0}=q(t),\;Q_{1}=q^{(1)}(t),....,%
\;Q_{k-1}=q^{(k-1)}(t)\;$together with $k\;$momenta
\begin{equation}
P_{i}(t)=\frac{\partial L}{\partial q^{(i+1)}} +
\sum_{j=1}^{k-i-1}\left(-\frac{d}{dt}  \right)^j \frac{\partial
L}{\partial q^{(j+1)}}, \quad i=0, \dots, (k-1). \label{GENMOM}
\end{equation}%
In the sequel we avoid writing the explicit time dependence
in the fields. The equation of motion for  $q\;$will be of order
$2k$ requiring the fixing of $2k\;$initial conditions, which is
consistent with the existence of  $2k\;$degrees  of
freedom in phase space. The Hamiltonian is%
\begin{equation}
H\left( Q_{0},...,Q_{k-1};P_{0},...,P_{k-1}\right)
=\sum_{i=0}^{k-1}\;P_{i}Q_{i}-\;L(Q_{0},...,Q_{k-1},%
\;q^{(k)}(P_{k-1},Q_{0},...,Q_{k-1}))  \label{HAMHOTD}
\end{equation}
where the non-degeneracy assumption implies that
we can solve $q^{(k)}\;$as a function of
$Q_{0},...,Q_{k-1},P_{k-1}.\;$The above expression is linear in the
momenta $P_{i},\;i=0,...,k-2$, thus making the Hamiltonian
(\ref{HAMHOTD}) unbounded from below, independently of the
interaction terms included in the Lagrangian.

Since we are interested in dealing with HOTD corrections in the
action as perturbations upon standard theories, we must rely on a
perturbation procedure which (i) retains the original number of
degrees of freedom and (ii) produces a free Hamiltonian which is bounded from
below as an adequate starting point for quantization. Such
a method has been already developed in Refs. \cite{ELIEZER},\ \cite%
{CHENGETAL} and we present here a brief summary of it. To point
out some of its  basic features let us consider the simplest
framework of a Lagrangian depending upon accelerations
$L(q,\dot{q},\ddot{q})$, where the HOTD contribution arising from
$\frac{1}{2}\, g \, \ddot{q}^2$ is only present as a perturbation
characterized by a small parameter $g$
 \be
L(q,\dot{q},\ddot{q})=L_0(q,\dot{q})+ \frac{1}{2}\, g \, \ddot{q}^2.
\ee
The standard procedure of extremizing the action leads to
\begin{equation}
\delta S=\delta \int_{t_1}^{t_2} dt \, L(q, \dot{q}, \ddot{q})=
\left[P_0 \delta q +P_1 \delta
\dot{q}\right]_{t_1}^{t_2}+\int_{t_1}^{t_2} dt E\left(q,{\dot q},
\ddot{q}, q^{(4)}\right)\delta q(t), \label{VARACT} \end{equation}
where \be E(q, \, \dot{q},\, \ddot{q},\, q^{(4)})=0,
\label{EQSMOT}\ee is the fourth order  equation of motion.  From
Eq. (\ref{VARACT}) the momenta can be directly read off as
 \be P_0= \frac{\partial L_0}{\partial \dot{q}}-g q^{(3)},
 \qquad P_1=g \ddot{q}.  \ee
From the simple form assumed for the HTOD term, which is appropriate for the non-degenerate situation,  we can solve for  $ \ddot{q}$  in terms of
 $P_1$ and $q^{(3)}$ in terms of $P_0$. Nevertheless, notice that
 both substitutions carry the non-analytical  factor $1/g$. This
 is precisely what makes non-trivial a perturbative expansion around
 $g=0$.

 The Hamiltonian $H$ and symplectic form $\Omega$ are defined according to the Ostrogradski
 method as
\begin{equation}H(q,\, \dot{q},\, P_0,\, P_1)=P_0
\dot{q}+P_1 \ddot{q}- L, \qquad \Omega = dP_0 \wedge d {q}+ dP_1
\wedge d {\dot q}.
\end{equation}
The dangerous contributions to the Hamiltonian arise from the
non-analytic term $P_1^2/2g$ together with the unbounded piece
$P_0\, \dot{q}$.

Let us summarize now the general perturbation scheme for the non-degenerate
case developed in Ref.\cite{CHENGETAL}, which
applies to systems of the general form \be L=L_0(q \, , \,
\dot{q})+g L_1(q \, , \, \dot{q}\, , \, \dots \, , \,  q^{(n)}).
\ee
In order to obtain the
appropriate Hamiltonian to order $g^k$, one starts by solving the
equations of motion to order $g^{(k-1)}$ under the requirement of
expressing all higher order derivatives $q^{(n)} , \,\,  n\geq 2$
only in terms of $ q$ and $ \dot{q}$, which provides the  first
step to define the effective variables of the problem. This will
allow us to rewrite \be H=H(q, \, \dot{q}),\qquad \Omega=
\omega(q, \, \dot{q})\, d\dot{q}\, \wedge\, dq , \ee from where we
can read the bracket $\{ q \, , \, \dot{q}\}$. Next we look for an
invertible change of variables $Q(q, \, \dot{q}),\, P(q, \,
\dot{q})$ such that \be 1+O(g^{k+1})= \{Q\, , \,
P\}=\left(\frac{\partial Q}{\partial q}\frac{\partial P}{\partial
\dot{q}}- \frac{\partial Q}{\partial \dot{q}}\frac{\partial
P}{\partial q}\right)\{ q \, , \, \dot{q}\}.\ee Finally,  the
Hamiltonian ${\tilde H}(Q,\,P)=H(q(Q,P)\, , \, \dot{q}(Q,P))$
together with the Poisson bracket $\{Q , P\}=1$ define the
physical effective system to the order considered. This
Hamiltonian will be bounded from below provided that the initial one
obtained from $L_0$ is. The effective Lagrangian is given by
${\tilde L}(Q\, , \dot{Q})= P\, \dot{Q}-{\tilde H}$ and the
corresponding Euler-Lagrange equations reproduce those of the
original system to the order considered. Since this Lagrangian is
first order in the time derivatives, the quantization is straightforward.  A formal proof
of self-consistency to all orders is provided in
Ref.\cite{CHENGETAL}. As clearly shown in one of the examples of
such reference,  the physical meaning of reducing the original
phase space to that generated by $q$ and $\dot{q}$ in the
perturbative formulation is to suppress the  excitation of high
energy modes in a way consistent with the exact evolution and only
to allow further  excitations of the low energy modes already
present in the zeroth order system. The method has been
generalized to field theory in Ref. \cite{CHENGETAL1}.

\section{The complex scalar field}

Before applying the method of Ref. \cite{CHENGETAL}  to our problem we need to identify  the reduced phase
space of this sector of the model which arises due to the presence of constraints, as it is shown in the
next subsection.
\subsection{Reduced phase space dynamics}
We  start from
\begin{equation}
\mathcal{L}_{scalar}(\phi ^{\ast },\phi )=-\phi ^{\ast }\left( \partial
^{2}+m^{2}\right) \phi +ig\phi ^{\ast }\left( n^{\mu }\partial _{\mu
}\right) ^{3}\phi ,\;  \label{FREESCALAR}
\end{equation}%
which describes the charged scalar field extension of the original MP
Lagrangian. \ In the rest frame $n^{\mu }=(1,\mathbf{0})\;$the equations of
motion are
\begin{equation}
\ddot{\phi}=\nabla ^{2}\phi -m^{2}\phi +ig\phi ^{(3)},  \label{EQMOS}
\end{equation}%
together with its complex conjugate, which exhibit the HOTD character of the
model. Inserting a plane wave in the above equation provides the MDR
as  solutions of a cubic equation. The condition for recovering the standard energy-momentum
relation in the limit $g\rightarrow 0$ requires that  this cubic equation has three real solutions, one of which will be non-analytical in $g$, while the remaining two will be recovered in the perturbative approach to be used in the sequel. The above requirement demands
\be
g^2E^2_k < \frac{4}{27}, \qquad E_k=\sqrt{|\mathbf{k}|^2+m^2}, \label{MOMBOUND}
\ee
which sets an upper limit to $|\mathbf{k}|$ in a way analogous to the photonic and fermionic cases. For future comparison we write here the MDR  obtained directly by solving  the cubic equation to order $g^2$
\be
\omega _{\pm}(\mathbf{k}) =\pm\left[E_k \pm \frac{1}{2}\allowbreak gE_k^{2}+\frac{5}{8}g^{2}E_k^{3}\right].
\label{EXMDR}
\ee

The coordinate space variables  of the model are
\begin{equation}
\phi ^{\ast },\phi ,\;\dot{\phi}^{\ast },\dot{\phi},  \label{COORDSP}
\end{equation}%
and our notation is
\be
\phi ^{(n)}=\frac{\partial ^{n}\phi }{\partial t^{n}},\;\;\phi ^{(1)}=\dot{%
\phi},\;\;\phi ^{(2)}=\ddot{\phi}\;.
\ee
The presence of third-order time derivatives in the equation of
motion (\ref{EQMOS}), instead of fourth order ones, precludes the existence of
constraints in the Hamiltonian formalism. In order to deal with this
construction we have found it convenient to perform one integration
by parts in the dimension five operator appearing in Eq.
(\ref{FREESCALAR}) . This leads to
\begin{equation}
\mathcal{L}=\dot{\phi}^{\ast }\dot{\phi}+\phi ^{\ast }(\nabla
^{2}-m^{2})\phi -ig\dot{\phi}^{\ast }\ddot{\phi}.
\end{equation}%
The canonically conjugated momenta corresponding to the coordinates\ (\ref%
{COORDSP}) are
\begin{eqnarray}
\Pi _{\phi ^{\ast }} &=&\dot{\phi}-\;ig\ddot{\phi},\;\;\;\Pi _{\phi }=\dot{%
\phi}^{\ast }+ig\ddot{\phi}^{\ast },\;\;  \label{MOM1} \\
\Pi _{\dot{\phi}} &=&-ig\dot{\phi}^{\ast },\;\;\;\;\;\;\Pi _{\dot{\phi}%
^{\ast }}=0,  \label{MOM2}
\end{eqnarray}%
respectively. The Eqs.\ (\ref{MOM2}) are  primary constraints and the extended
Hamiltonian density is%
\be
\mathcal{H}_{E} =\Pi _{\phi }{\dot{\phi}}-\frac{i}{g}{\dot{\phi}}\Pi _{%
\dot{\phi}}-\phi ^{\ast }(\nabla ^{2}-m^{2})\phi +\frac{i}{g}\Pi _{\phi
^{\ast }}\Pi _{\dot{\phi}} + \alpha \Pi _{{\dot{\phi}}^{\ast }}+\beta \left( \Pi _{\dot{\phi}}+ig\dot{%
\phi}^{\ast }\right).  \label{EXTHAM}
\ee
The evolution of the primary constraints fixes the Lagrange multipliers\ $%
\alpha ,\beta $, so that no additional constraints arise and the original
ones are second class. The number of coordinate degrees of freedom per space
point is then $\frac{1}{2}(2\times 4-2)=3\;$which exceeds the usual two 
associated with the standard charged scalar field. We impose the constraints
\begin{equation}
\chi _{1}=\Pi _{{\dot{\phi}}^{\ast }},\;\;\chi _{2}=\;\Pi _{\dot{\phi}}+ig%
\dot{\phi}^{\ast }, \;  \label{CONSTR}
\end{equation}%
strongly by introducing the corresponding Dirac brackets and verify that
the resulting brackets for the reduced space variables
\begin{equation}
(\phi ,\Pi _{\phi }),\;\;(\phi ^{\ast },\Pi _{\phi ^{\ast }}),\;\;(\dot{\phi}%
,\Pi _{\dot{\phi}}),\;  \label{REDSPACE}
\end{equation}%
remain unchanged with respect to the original Poisson brackets, thus
preserving the original symplectic structure. The reduced phase space
Hamiltonian density is then%
\begin{equation}
\mathcal{H}_{R}=\Pi _{\phi }{\dot{\phi}}+\frac{i}{g}\Pi _{\dot{\phi}}\left(
\Pi _{\phi ^{\ast }}-{\dot{\phi}}\right) -\phi ^{\ast }(\nabla
^{2}-m^{2})\phi ,  \label{REDHAM}
\end{equation}%
which exhibits a non-positive definite character due to  terms linear in the
coordinates and momenta; as well as the  non-analytic behavior in $1/g$
 that prevents  naive attempts of constructing perturbative schemes around $g=0$.
From Eq. (\ref{REDHAM}) we  obtain the equations of motion%
\begin{eqnarray}
\dot{\Pi}_{\phi } &=&(\nabla ^{2}-m^{2})\phi ^{\ast },\quad \dot{\Pi}_{\phi
^{\ast }}=(\nabla ^{2}-m^{2})\phi ,\quad \dot{\Pi}_{\dot{\phi}}=-\left( \Pi
_{\phi }-\frac{i}{g}\Pi _{\dot{\phi}}\right) ,\;\;  \label{EQMOTMOM} \\
\dot{\phi}^{\ast } &=&\frac{i}{g}\Pi _{\dot{\phi}},\quad\;\frac{\partial \dot{%
\phi}}{\partial t}=\ddot{\phi}=\frac{i}{g}\left( \Pi _{\phi ^{\ast }}-{\dot{%
\phi}}\right) .  \label{EQMOTCOORD}
\end{eqnarray}

From them it is a direct matter to recover the Lagrangian equation (\ref%
{EQMOS}) and, consistently in an independent way, the corresponding complex
conjugate expression.

\subsection{The perturbative expansion}

\subsubsection{The symplectic form}

Regarding the symplectic structure of the theory
let us indicate our conventions for the field theory case.
Denoting generically $Z^{A}(t,\mathbf{x}%
)$ the bosonic phase space fields (both coordinates and momenta), the Poisson brackets relations%
\begin{equation}
\left\{ Z^{A}(t,\mathbf{x}),\;Z^{B}(t,\mathbf{x}^{\prime })\right\}
=W^{AB}(t;\mathbf{x,x}^{\prime }),
\end{equation}%
are encoded in the corresponding symplectic two form $\Omega (t)$ through
the relation
\be
\Omega (t)=\frac{1}{2}\int d^{3}\mathbf{x\;}d^{3}\mathbf{x}^{\prime
}\;W_{AB}(t;\mathbf{x,x}^{\prime })\;dZ^{A}(t,\mathbf{x})\wedge dZ^{B}(t,%
\mathbf{x}^{\prime }),
\ee
where $W_{AB}(t;\mathbf{x,x}^{\prime })=-W_{BA}(t;\mathbf{x}^{\prime }%
\mathbf{,x}),$ which can be considered as an antisymmetric matrix in its
discrete as well as continuous indices, is such that%
\be
\int d^{3}\mathbf{x}^{\prime }\;W_{AB}(t;\mathbf{x,x}^{\prime })W^{BC}(t;%
\mathbf{x}^{\prime },\mathbf{x}^{\prime \prime })=\delta _{A}^{C}\;\delta
^{3}(\mathbf{x-x}^{\prime \prime }).
\ee
Our next step in the perturbative construction is to use the equations of
motion to second order in $g\;$in order to rewrite the exact symplectic form%
\begin{equation}
\Omega =\int d^{3}\mathbf{x}\;(d\Pi _{\phi }\wedge
d{{\phi}\;+\;}d\Pi
_{\phi ^{\ast }}\wedge d{{\phi}}^{\ast }+d\Pi _{{\dot{\phi}}}\wedge d{%
\dot{\phi})},
\end{equation}%
in terms only of the unperturbed variables $\phi ,\;\phi ^{\ast }$ together
with their first order time derivatives.

The approximation to order $g^{2}\;$ leads to%
\begin{equation}
\ddot{\phi}=\left( -\hat{E}^{2}\right) \left[ \left( 1+g^{2}\hat{E}%
^{2}\right) \phi +ig\dot{\phi}\right] +O(g^{3}),
\end{equation}%
together with it complex conjugate, which we also substitute in the
corresponding momentum expressions.
Here we have introduced the notation
\begin{equation}
\hat{E}^{2}=\left( m^{2}-\nabla _{\mathbf{x}}^{2}\right).
\end{equation}%
In this way the Hamiltonian density is%
\begin{equation}
\mathcal{H}={\dot{\phi}}^{\ast }\left( 1-2g^{2}\hat{E}^{2}\right) {\dot{\phi}%
}+ig\dot{\phi}^{\ast }\hat{E}^{2}\phi -ig\phi ^{\ast }\hat{E}^{2}{\dot{\phi}}%
+\phi ^{\ast }\hat{E}^{2}\phi,  \label{HAMDEN0}
\end{equation}
together with  the symplectic  form
\begin{eqnarray}
\Omega &=&\int d^{3}\mathbf{x}\left( -2igd\phi ^{\ast }\hat{E}^{2}\wedge d{%
\phi }+d{\dot{\phi}}^{\ast }\wedge \left( 1-g^{2}\hat{E}^{2}\right) d{\phi }%
\right)  \nonumber \\
&&+ \int d^{3}\mathbf{x}\left( d{\dot{\phi}}\wedge \left( 1-g^{2}\hat{E}%
^{2}\right) d{\phi }^{\ast }-igd\dot{\phi}^{\ast }\wedge d{\dot{\phi}}\right).
\label{SYMFORM}
\end{eqnarray}%
From Eq. (\ref{SYMFORM}) we read
\begin{equation}
W_{AB}(\mathbf{x,x}^{\prime })=\left[
\begin{array}{cccc}
0 & +2ig\hat{E}^{2} & 0 & -\left( 1-g^{2}\hat{E}^{2}\right) \\
-2ig\hat{E}^{2} & 0 & -\left( 1-g^{2}\hat{E}^{2}\right) & 0 \\
0 & \left( 1-g^{2}\hat{E}^{2}\right) & 0 & ig \\
\left( 1-g^{2}\hat{E}^{2}\right) & 0 & -ig & 0%
\end{array}%
\right] \delta ^{3}(\mathbf{x}-\mathbf{x}^{\prime }).  \label{ORIGPB}
\end{equation}%
The notation here is $A,B=1,2,3,4, \,\,$ with $Z^{1}=\phi ,\; Z^{2}=\phi ^{\ast
}, \; Z^{3}=\dot{\phi}\;$ and $Z^{4}=\dot{\phi}^{\ast }$.\ Inverting the matrix
(\ref{ORIGPB})  to second order in $g$ we obtain%
\be
W^{AB}(\mathbf{x,x}^{\prime })=\left[
\begin{array}{cccc}
0 & -ig & 0 & 1+3g^{2}\hat{E}^{2} \\
ig & 0 & 1+3g^{2}\hat{E}^{2} & 0 \\
0 & -\left( 1+3g^{2}\hat{E}^{2}\right) & 0 & -2ig\hat{E}^{2} \\
-\left( 1+3g^{2}\hat{E}^{2}\right) & 0 & 2ig\hat{E}^{2} & 0%
\end{array}%
\right] \delta ^{3}(\mathbf{x}-\mathbf{x}^{\prime }),
\ee
from where we read the non-zero brackets among the fields and their first order time derivatives.
\begin{eqnarray}
\left\{ \phi \left( \mathbf{x,}t\right) ,\phi ^{\ast }\left( \mathbf{x}%
^{\prime },t\right) \right\} &=&-ig\delta ^{3}(\mathbf{x}-\mathbf{x}^{\prime
}),\; \\
\left\{ \phi \left( \mathbf{x,}t\right) ,\dot{\phi}^{*}\left( \mathbf{x}^{\prime
},t\right) \right\} &=&\left( 1+3g^{2}\hat{E}^{2}\right) \delta ^{3}(\mathbf{x%
}-\mathbf{x}^{\prime }),\;\; \\
\;\left\{ \phi ^{\ast }\left( \mathbf{x,}t\right) ,\dot{\phi}\left( \mathbf{x%
}^{\prime },t\right) \right\} &=&\left( 1+3g^{2}\hat{E}^{2}\right)
\delta
^{3}(\mathbf{x}-\mathbf{x}^{\prime }),\;\;\; \\
\left\{ \dot{\phi}\left( \mathbf{x,}t\right) ,\dot{\phi}^{\ast }\left(
\mathbf{x}^{\prime },t\right) \right\} &=&-2ig\hat{E}^{2}\delta ^{3}(\mathbf{%
x}-\mathbf{x}^{\prime }). \label{NEWPB}
\end{eqnarray}%
\subsubsection{New canonical variables}
The next step in the procedure is to introduce new canonical variables
having the standard symplectic form in such a way that the final
Hamiltonian density has the quadratic term in the momenta
normalized to one. To this end we define, to order
$g^{2}$, the following new coordinates
$\tilde{\phi},\tilde{\phi}^{\ast }$
and momenta\ $\Pi _{\tilde{\phi}},\;\Pi _{\tilde{\phi}^{\ast }}$%
\be
\tilde{\phi}=\left( P\phi -\frac{ig}{2}Q\dot{\phi}\right) ,\;\;\Pi _{\tilde{%
\phi}}=\left( R\dot{\phi}^{\ast }-ig\hat{E}^{2}S\phi ^{\ast }\right)
,\;\;\;\;
\ee
together with the corresponding complex conjugates: $\tilde{\phi}^{\ast
}=\left( \tilde{\phi}\right) ^{\ast },\;\Pi _{\tilde{\phi}^{\ast }}=(\Pi _{%
\tilde{\phi}})^{\ast }.\;$The imposition of standard Poisson brackets among
the tilde variables leads to the conditions%
\begin{equation}
P=Q,\quad R=S,\quad PR=1-\frac{3g^{2}}{2}\hat{E}^{2}. \label{COND1}
\end{equation}%
The inverse functions are given by
\begin{eqnarray}
\phi  &=&\left( 1+\frac{1}{2}g^{2}\hat{E}^{2}\right) \left( \frac{1}{P}%
\tilde{\phi}+\;\frac{1}{R}\frac{ig}{2}\Pi _{\tilde{\phi}^{\ast }}\right), \\
\dot{\phi} &=&\left( 1+\frac{1}{2}g^{2}\hat{E}^{2}\right) \left( \frac{1}{R}%
\Pi _{\tilde{\phi}^{\ast }}-\frac{1}{P}ig\hat{E}^{2}\tilde{\phi}\right).
\end{eqnarray}
Substituting in the
expression (\ref{HAMDEN0}) for the Hamiltonian density we obtain, to order $g^2$,
\begin{equation}
\mathcal{H}=\Pi _{\tilde{\phi}}\Pi _{\tilde{\phi}^{\ast }}+\frac{ig}{2}%
\left( \tilde{\phi}^{\ast }\hat{E}^{2}\Pi _{\tilde{\phi}^{\ast }}-\Pi _{%
\tilde{\phi}}\hat{E}^{2}\tilde{\phi}\right) +\tilde{\phi}^{\ast }\left(
\hat{E}^{2}+\frac{5g^{2}}{4}\left( \hat{E}^{2}\right) ^{2}\right) \tilde{\phi%
},  \label{HAMDEN1}
\end{equation}%
where the momenta are given by
\be
\Pi _{\tilde{\phi}^{\ast }} =\frac{\partial \tilde{\phi}}{\partial t}+\frac{ig}{2}\hat{E}%
^{2}\tilde{\phi}, \quad
\Pi _{\tilde{\phi}} =\frac{\partial \tilde{\phi}^{\dag }}{\partial t}-\frac{ig}{2}\hat{E}%
^{2}\tilde{\phi}^{\dag }. \label{MOM}
\ee
The Hamiltonian density can be conveniently written as
\be
\mathcal{H}= \frac{\partial \tilde{\phi}^{\dag }}{\partial t} \frac{\partial \tilde{\phi}}{\partial t%
} +\tilde{\phi}^{\dag }\left( \hat{E}^{2}+g^{2}\left(
\hat{E}^{2}\right) ^{2}\right) \tilde{\phi}. \label{CONVHAM}
\ee

\subsubsection{The modified dispersion relations}

Finally we  construct the corresponding effective Lagrangian density
\be
L =\frac{\partial {\tilde{\phi}^{\ast }}}{\partial t}\frac{\partial {\tilde{\phi}}}{\partial t}+\frac{ig}{2}%
\left( \frac{\partial {\tilde{\phi}^{\ast }}}{\partial t}\hat{E}^{2}\tilde{\phi}-\frac{\partial {\tilde{\phi}}}{\partial t}%
\hat{E}^{2}\tilde{\phi}^{\ast }\right) -\tilde{\phi}^{\ast }\left( \hat{E}^{2}+g^{2}\left( \hat{E}^{2}\right)
^{2}\right) \tilde{\phi},
\ee
leading to the equation of motion 
\begin{equation}
\frac{\partial^{2}{\tilde{\phi}}(g)}{\partial t^{2}}+ig\hat{E}^{2}\frac{\partial {\tilde{\phi}}(g)}{\partial t}
+\left( \hat{E}^{2}+g^{2}\left( \hat{E}^{2}\right) ^{2}\right) \tilde{\phi%
}(g)=0. \label{EQMOT}
\end{equation}
Using a plane wave type solution one finds the MDR
for positive frequencies to be
\be
\omega _{1} \left( \mathbf{k}\right)=E_{k}\left( 1+\frac{1}{2}gE_{k}+\frac{5}{8}%
(gE_{k})^{2}\right) , \quad
\omega _{2}\left( \mathbf{k}\right)=E_{k}\left( 1-\frac{1}{2}gE_{k}+\frac{5}{8}%
(gE_{k})^{2}\right). \label{PERTMDR}
\ee
 The  energy dependence in the corresponding plane wave are $e^{-i\omega_1t}$ and $e^{i\omega_2t}$  respectively. Let us emphasize two points at this level: (i) the MDR (\ref{PERTMDR}) coincide with those derived from the exact equation in (\ref{EXMDR}).   (ii) the theory is charge
conjugation invariant, which means that the field $\phi(g)$ together with the charge conjugate field
$\phi_C(g)=\phi^{*}(-g)$ are solutions of the equation of motion (\ref{EQMOT}), as can be directly verified.

\subsection{The quantization}
It proceeds along the standard lines of the complex scalar field, the only
difference being that the MDR are given by Eq.(\ref{PERTMDR}). In fact, we introduce the independent set
of creation-annihilation operators%
\begin{equation}
\left[ a(\mathbf{k}),\;a^{\dag }(\mathbf{k}^{\prime })\right] =\delta ^{3}(%
\mathbf{k-k}^{\prime }),\;\;\left[ b(\mathbf{k}),\;b^{\dag }(\mathbf{k}%
^{\prime })\right] =\delta ^{3}(\mathbf{k-k}^{\prime }),  \label{CAOPCR}
\end{equation}%
with standard vacuum
\begin{equation}
a(\mathbf{k})|0\rangle =0=b(\mathbf{k})|0\rangle.
\end{equation}%
In terms of these operators, we expand the complex field $\tilde{\phi}\;$
as%
\begin{equation}
\tilde{\phi}(x)=\int \frac{d^{3}\mathbf{k}}{\sqrt{(2\pi )^{3}}}\,\frac{1%
}{\sqrt{2\Omega (\mathbf{|k|)}}}\,\left( a(\mathbf{k})e^{-i\left( \omega_1t-%
\mathbf{k}\cdot \mathbf{x}\right) }+b^{\dag }(\mathbf{k})e^{i\left( \omega
_2t-\mathbf{k}\cdot \mathbf{x}\right) }\right),
\end{equation}%
together with its hermitian conjugate. The normalization factor $\Omega (\mathbf{|k|)}$ is determined by demanding
the equal time commutation relation
\be
\left[\tilde{\phi}(\mathbf{x}), \Pi_{\tilde{\phi}}(\mathbf{x'}) \right]=i\delta^{3}(\mathbf{x}-\mathbf{x'}),
\ee
and it is given by
\be
\Omega (\mathbf{|k|)}=E_{k}\left( 1+\frac{5}{8}%
g^{2}E_{k}^{2}\right). \label{NORMF}
\ee
The remaining canonical commutation relations among the field operators can  also be recovered,
in virtue of  the relations (\ref{CAOPCR}) and (\ref{NORMF}) . Starting from (\ref{CONVHAM}), the normal ordered Hamiltonian turns out to be
\begin{equation}
H=\int d^{3}\mathbf{k}\;\left[ \omega _1\,a^{\dag }(\mathbf{k})a(\mathbf{k}%
)+\omega _2\, b^{\dag }(\mathbf{k})b(\mathbf{k})\right] .
\end{equation}%
This result can be considered as an additional  consistency check of the choice (\ref{NORMF}) to order $g^2$.
The interpretation of
$a^{\dag }(\mathbf{k}),a(\mathbf{k})$ ($b^{\dag }(\mathbf{k}),b(\mathbf{k})$%
) as creation and annihilation operators for positively (negatively) charged
particles also follows directly. In fact, the normal ordered charge operator is given by%
\begin{equation}
Q=\int d^{3}\mathbf{k}\;\;\left[ a^{\dag }(\mathbf{k})a(\mathbf{k})-b^{\dag
}(\mathbf{k})b(\mathbf{k})\right].
\end{equation}%
The Feynman propagator is
\begin{equation}
i\Delta _{F}(x-y)=\langle 0|T({\tilde{\phi}}(x){\tilde{\phi}}^{\dag }(y))|0\rangle ,
\end{equation}%
which can be  separated into retarded and advanced  pieces
\begin{equation}
i\Delta _{F}(x-y)=\theta (x_{0}-y_{0})\langle 0|{\tilde{\phi}}(x){\tilde{\phi }}^{\dag
}(y)|0\rangle +\theta (y_{0}-x_{0})\langle 0|\tilde{\phi}^{\dag }(y)\tilde{\phi}
(x)|0\rangle.
\end{equation}%
The standard calculation yields
\begin{equation}
\Delta _{F}(x-y)=\int \frac{d^{4}k}{(2\pi )^{4}}\,\frac{e^{-ik(x-y)}}{%
{k_{0}^{2}-k_{0}gE_{k}^{2}-E_{k}^{2}%
\left( 1+g^{2}E_{k}^{2}\right) +i\epsilon } },
\end{equation}%
where the denominator reproduces  exactly  the  momentum space version of  the effective equation of motion 
 (\ref{EQMOT}).

\begin{theacknowledgments}
  LU thanks the organizers of the Third Mexican Meeting on Mathematical and Experimental Physics for their invitation. The work of  CMR, JDV and LU has been  partially supported by the
  projects  DGAPA-UNAM  \# IN109017 and CONACYT \# 55310. JDV also acknowledges partial support from the project CONACYT \# 47211.
\end{theacknowledgments}

\end{document}